\RequirePackage{fix-cm}
\documentclass[twocolumn,epjc3]{svjour3}
\smartqed  
\RequirePackage{graphicx}
\journalname{Eur. Phys. J. C}
\usepackage{amsmath}
\usepackage{hyperref}
\usepackage{cite}

\begin{document}


\title{Stochastic quantum inflation for a canonical scalar field with linear self-interaction potential}

\author{Grigoris Panotopoulos \thanksref{e1,addr1}}

\thankstext{e1}{e-mail: grigorios.panotopoulos@tecnico.ulisboa.pt}


\institute{CENTRA, Instituto Superior T{\'e}cnico, Universidade de Lisboa,\\ Av. Rovisco Pais 1, Lisboa, Portugal\label{addr1}}


\date{Received: date / Accepted: date}

\maketitle

\begin{abstract}
We apply Starobinsky's formalism of stochastic inflation to the case of a massless minimally coupled scalar field with linear self-interaction
potential. We solve the corresponding Fokker-Planck equation exactly, and obtain analytical expressions for the
stochastic expectation values.

\PACS{PACS code1 \and PACS code2 \and more}
\end{abstract}

\section{Introduction}

Inflation \cite{starobinsky1,inflation} is widely accepted as the standard paradigm of the early Universe. The first reason is due to the fact that several long-standing puzzles of the Hot Big-Bang model, such as the horizon, flatness, and monopole problems, find a natural explanation in the framework of inflationary Universe. In addition, and perhaps the most intriguing feature of inflation, is that it gives us a causal interpretation of the origin of the Cosmic Microwave Background (CMB) temperature anisotropies \cite{Abazajian:2013vfg}, while at the same time it provides us with a mechanism to explain the Large-Scale Structure (LSS) of the Universe, since quantum fluctuations during the inflationary era may give rise to the 
primordial density perturbations \cite{mukhanov}.

Despite its success a theory of inflation is still missing, since as of today we do not know neither what the inflaton is nor why the inflaton potential is so flat. Furthermore, the inflaton potential cannot be derived from a fundamental theory in a unique way. All we have is a large collection of inflationary models (see e.g.\cite{models}, and for a classification of inflationary models see e.g. \cite{Kinney:2003uw}). In single-field inflationary models with a canonical scalar field in Einstein's General Relativity (GR) there are basically four classes of models, namely large field models, small field models, hybrid models and linear inflation that lies between large and small field models. After the latest Planck results \cite{planck2015} many inflationary models based on monomial potentials have been ruled out or are disfavored by data. There are of course the Starobinsky's model \cite{starobinsky1} as well as the Higgs inflation model \cite{shapo} which are in agreement with the data, but they are extensions of GR. The linear inflaton potential is still in agreement with observations, but difficult to realize in particle physics models. However, recently it was shown that the Coleman-Weinberg potential \cite{CW} together with a nonminimal coupling of the inflaton to gravity leads to attractor solutions that interpolate between quadratic inflation, which is ruled out by recent measurements, and linear inflation, which lies within the allowed region \cite{paper1,paper2}. In particular, first in \cite{paper1} it was shown that the predictions of linear inflation for the observables can be achieved in the context of well-defined quantum field theory, without introducing complicated interactions by hand. Then in \cite{paper2} the authors extended the previous discussion, presenting a more detailed study of the parameter space, and they also added a discussion on reheating.

During inflation infrared logarithms arise in the expectation values of operators of quantum field theories that contain massless minimally coupled scalar fields
or gravitons. For an incomplete list see e.g. \cite{calculations} and references therein. These terms are powers of the logarithm of the inflationary scale factor
$a=exp(H t)$, with $H$ being the Hubble constant, and are very exciting because they may signal important quantum effects \cite{QE}. However, their continued growth implies that when inflation has proceeded for a long time, the large logarithms eventually overcome the small coupling constants. Therefore the effects become non-perturbative and perturbation theory breaks down. A natural approach to obtain non-perturbative information
is the leading logarithm approximation \cite{woodard1}, in which one attempts to sum the series comprised of just the leading infrared logarithms at each order. For scalar fields with non-derivative interactions Starobinsky's technique of stochastic inflation \cite{starobinsky2,Starobinsky:1994bd} recovers the leading infrared logarithms at each order, and the series of these leading effects at all orders can be resummed \cite{resum} to give non-perturbative predictions.

Given the observational bound on the tensor-to-scalar ratio $r < 0.07$ \cite{Array:2015xqh}, the slow-roll parameter $\epsilon=r/16$ is extremely close to zero, $\epsilon < 0.0044$, and therefore the de Sitter spacetime is an excellent approximation to inflation. The goal of this work is to study the effect of a minimally coupled scalar field with a linear self-interaction potential using Starobinsky's formalism.
Our work is organized as follows: After this introduction, we briefly summarize stochastic processes in the Langevin and Fokker-Planck approach in the second section. Then we apply this formalism to the case of a canonical scalar field with a linear potential following the Starobinsky's technique in section three. 
Finally we conclude in the last section.

\section{Stochastic processes}

The observation that small pollen grains, when suspended in water, are found to be in a very animated and irregular state of motion, was first systematically investigated by the Scottish botanist Robert Brown in 1827, and the observed phenomenon naturally was named after him. First A. Einstein \cite{einstein} in 1905 and independently M. Smoluchowski in 1906 \cite{mariam}, and some time later Paul Langevin \cite{langevin} in 1908 (see \cite{english} for a translation of the original Langevin paper in English) explained Brownian motion using different but equally successful approaches. From the one hand, Einstein's analysis was based on the diffusion equation
\begin{equation}
\frac{\partial f(t,x)}{\partial t}=D \frac{\partial^2 f(t,x)}{\partial^2 x}
\end{equation}
where D is the diffusion coefficient, with the initial condition $f(x,t=0)=\delta(x)$ where $\delta(x)$ is Dirac's delta function. The solution of the diffusion equation is given by \cite{arfken}
\begin{equation}
f(t,x)=\frac{1}{\sqrt{4 \pi D t}} exp(-\frac{x^2}{4Dt})
\end{equation}
and the mean of the square of displacement is given by $\langle x^2 \rangle = 2 D t$.
On the other hand, Langevin started from Newton's equation of motion assuming a Stokes's drag force and a random thermal force due to the continuous bombardment from the molecules of the liquid. Although he did not exploit all richness of his model, Langevin obtained in the late-time regime Einstein's result, namely $\langle x^2 \rangle = 2 (k_B T/6 \pi a \mu) t$, where $k_B$ is Boltzmann's constant, T is the temperature, $\mu$ is the fluid viscosity, and
$a$ is the radius of the particle. Therefore, thanks to Langevin's approach the Brownian motion, diffusion as well as the random walk were all linked together, a view that soon quantified experimentally by Perrin \cite{perrin}. Therefore the diffusion coefficient can be computed in terms of properties of the fluid and the Brownian particles, which is the Einstein-Stokes formula $D=(k_B T)/(6 \pi a \mu)$.

In modern times Langevin's approach is still in use. The Langevin equation for the process $x(t)$ (let us call it
the position of a moving particle) reads
\begin{equation}
\dot{x}=A(x)+\xi(t)
\end{equation}
where $A(x)$ is an external applied force, the dot denotes derivative with respect to time, and $\xi(t)$ is assumed to be a Gaussian white noise
\begin{eqnarray}
\langle \xi(t) \rangle & = & 0 \\
\langle \xi(t_1) \xi(t_2) \rangle & = & 2 D \delta(t_1-t_2)
\end{eqnarray}
Since the random force $\xi(t)$ is not known, we can only compute mean values of powers of the position, or the moments, $\langle x^n \rangle$, once the density probability function is known. The density probability function $u(t,x)$ satisfies the corresponding Fokker-Planck (FP) equation \cite{textbook}
\begin{equation}
\frac{\partial u(t,x)}{\partial t}=-\frac{\partial (A(x) u(t,x))}{\partial x}+D \frac{\partial^2 u(t,x)}{\partial^2 x}
\end{equation}
where the first term is due to the external force while the second term is the diffusion term with a constant diffusion coefficient D. If we ignore the external applied force, the FP equation reduces to the standard diffusion equation. That explains why the Einstein's and Langevin's approaches were equally successful. Solving the FP equation we then can compute the moments performing the integrals
\begin{equation}
\langle x^n \rangle=\int_{-\infty}^{\infty} dw u(t,w) w^n
\end{equation}
and they are functions of time. Although seemingly quite different equations, when the diffusion equation and the FP equation at hand have the same number of symmetries, or equivalently when the following condition is satisfied \cite{symmetries}
\begin{equation}
2DA^\prime(x)+A(x)^2=c_0+c_1 x+c_2 x^2
\end{equation}
the FP equation can be recast into the diffusion equation, and therefore admits an exact analytical solution.

\section{Application of Starobinsky's formalism to a massless canonical scalar field with a linear potential}

In this section we apply Starobinsky's formalism \cite{starobinsky2,Starobinsky:1994bd} to a scalar field $\Psi$ described by the Lagrangian
\begin{equation}
\mathcal{L} = - \sqrt{-g} \left( \frac{1}{2} g^{\mu \nu} \partial_\mu \Psi  \partial_\nu \Psi + V(\Psi) \right)
\end{equation}
where $V(\Psi)=- M^3 \Psi$ is the self-interaction potential of the scalar field taken to be linear in $\Psi$, with $M$ being a mass scale.
The resulting Klein-Gordon equation reads
\begin{equation}
\ddot{\Psi} + 3 H \dot{\Psi} - \frac{\nabla^2 \Psi}{a^2} + V_{,\Psi} = 0
\end{equation}
where $V_{,\Psi}$ is the derivative of the scalar potential with respect to the scalar field. We take the onset of inflation to be at $t=0$,
and we work perturbatively around the non-dynamical de Sitter background
\begin{equation}
ds^2 = -dt^2 + exp(2 H t) d\vec{x} \: d\vec{x}
\end{equation}
with the inflationary scale factor being $a(t) = exp(Ht)$, and $H$ being the Hubble constant. 

In the leading logarithm approximation the scalar field behaves like a stochastic variable $\phi$ satisfying Langevin's equation
\begin{equation}
\dot{\phi}(t,\vec{x})=f(t,\vec{x})-\frac{V_{,\phi}}{3 H}
\end{equation}
where the stochastic source $f(t,\vec{x})$ has the properties of the Gaussian white noise
\begin{equation}
\langle f(t_1,\vec{x}) f(t_2, \vec{x}) \rangle = \frac{H^3}{4 \pi^2} \delta(t_1-t_2)
\end{equation}
(see however eq. (11) of \cite{Starobinsky:1994bd} for the general case where the two space-like points $\vec{x_1},\vec{x_2}$ are taken to 
be different). Langevin's equation above can be rigorously derived from the underlying quantum field theory \cite{starobinsky2}, or it can be easily 
obtained from the Klein-Gordon equation by applying the stochastic rules of \cite{woodard2}, namely 

- During inflation the scale factor varies faster than the scalar field, and therefore the Hubble friction term dominates over the spatial derivative and 
the second time derivative terms.

- We replace the full field $\Psi$ with its stochastic counterpart $\phi$.

- We make the substitution $\dot{\phi} \rightarrow \dot{\phi}-f$, where $f$ is the stochastic source.

Since the scalar potential is linear, the corresponding Fokker-Planck equation takes the form
\begin{equation}
\frac{\partial u(t,\phi)}{\partial t}=-\frac{M^3}{3H} \frac{\partial u(t,\phi)}{\partial \phi}+\frac{\Delta}{2} \frac{\partial^2 u(t,\phi)}{\partial^2 \phi}
\end{equation}
where we have put $\Delta=H^3/(4 \pi^2)$. It is trivial to check that the condition (8) is satisfied, and therefore
for this model we can find exact analytical solution of the FP equation.
According to the previous discussion, the FK equation can be recast in the diffusion equation $\omega_\tau=\omega_{yy}$ and the solution is given by
\begin{equation}
u(t,\phi)=h(t,\phi) \omega(\tau(t,\phi),y(t,\phi))
\end{equation}
where $h(t,\phi), y(t,\phi), \tau(t,\phi)$ are given by \cite{symmetries}
\begin{eqnarray}
h(t,\phi) & = & exp\left(\frac{M^3 \phi}{3H \Delta}-\frac{M^6 t}{18 \Delta H^2}\right) \\
y & = & \phi \\
\tau & = & \Delta t/2
\end{eqnarray}
Therefore, the final expression for the solution is given by
\begin{equation}
u(t,\phi)=\frac{1}{\sqrt{2 \pi \Delta t}} exp\left(-\frac{(\phi-\frac{M^3 t}{3H})^2}{2 \Delta t}\right)
\end{equation}
Therefore it is now straightforward to compute the first stochastic expectation values $\langle \phi \rangle$, $\langle \phi^2 \rangle$, $\langle \phi^3 \rangle$
and $\langle \phi^4 \rangle$, which are found to be
\begin{eqnarray}
\langle \phi \rangle & = & \frac{M^3 t}{3H} \\
\langle \phi^2 \rangle & = & \left( \frac{M^3 \: t}{3H} \right)^2+\Delta \: t \\
\langle \phi^3 \rangle & = & \left( \frac{M^3 \: t}{3H} \right)^3+\frac{M^3 \: t^2}{H} \Delta  \\
\langle \phi^4 \rangle & = & \left( \frac{M^3 \: t}{3H} \right)^4+6 \Delta \left( \frac{M^3}{3H} \right)^2 t^3+3 (\Delta \: t)^2
\end{eqnarray}
where we have made use of the Gaussian integrals
\begin{equation}
\int_{-\infty}^{\infty} dx \; exp(-\alpha x^2) = \frac{\sqrt{\pi}}{\sqrt{\alpha}}
\end{equation}
\begin{equation}
\int_{-\infty}^{\infty} dx \; x^2 \; exp(-\alpha x^2)  =  \frac{\sqrt{\pi}}{2 \alpha^{3/2}}
\end{equation}
\begin{equation}
\int_{-\infty}^{\infty} dx \; x^4 \; exp(-\alpha x^2)  =  \frac{3 \sqrt{\pi}}{4 \alpha^{5/2}}
\end{equation}
In the formulas obtained above, it is easy to check that when $M=0$ we recover the known result for the pure diffusion case.
Note that there are two contributions with different time dependence. In particular, the $M$ dependent terms are due to the classical
linear potential, while the $M$ independent terms are due to quantum effects, and they are just the ones of massless scalar fields.
Equivalently, we switch from the cosmological time $t$ to the inflationary scale factor $a$ by using $t=ln(a)/H$ and we find the following expressions
\begin{equation}
\langle \phi \rangle  =  \frac{M^3 ln(a)}{3 H^2}
\end{equation}
\begin{equation}
\langle \phi^2 \rangle  =  \frac{H^2 ln(a)}{4 \pi^2} \left( 1 + \frac{4 \pi^2 M^6}{9 H^6} ln(a) \right)
\end{equation}
\begin{equation}
\langle \phi^3 \rangle  =  \frac{M^3 ln^2(a)}{4 \pi^2} \left( 1 + \frac{4 \pi^2 M^6 ln(a)}{27 H^6} \right)
\end{equation}
\begin{equation}
\langle \phi^4 \rangle  =  \frac{3 H^4 ln^2(a)}{16 \pi^4} \left(1 + \frac{8 \pi^2 M^6}{9 H^6} ln(a) + \frac{16 \pi^4 M^{12}}{243 H^{12}} ln^2(a) \right)
\end{equation}
The factors $ln(a)$ associated to the $M$ independent quantum effects are the IR logarithms mentioned in the Introduction.
The model is characterized by two mass scales, $H,M$, and we can view the ratio $M/H$ as a dimensionless coupling constant. As already mentioned, 
even if $M/H$ is small, when inflation has proceeded for a long time the large logarithms eventually overcome the small coupling constant. 

Powers of $H \: t$ due to quantum effects of massless gravitons slow inflation \cite{Tsamis:1997za}. In this model, however, it is the classical 
effect of the scalar field rolling down the linear potential that slows inflation. As it has been shown in \cite{ref1,ref2}, 
massless canonical scalar fields (with vanishing potential)
contribute negligibly to the energy density in de Sitter, and lead to negligible backreaction. On the contrary, the classical homogeneous field rolling 
down the linear potential discussed here will eventually develop an energy density comparable to the cosmological constant.

Since the probability density function is Gaussian, it is characterized by only two parameters, and therefore all moments can be given in terms of the first two. However, in the following we shall compute the generic stochastic expectation value $\phi^{2n}$ for even powers and $\phi^{2n+1}$ for odd powers in closed form. 
To this end we change variable $z=\phi-(M^3t)/(3H)$, use the binomial expansion
\begin{equation}
(a+b)^m = \sum_{k=0}^m \frac{m!}{k! (m-k)!} a^{m-k} b^k
\end{equation}
and make use of the integral
\begin{equation}
\int_{-\infty}^{\infty} dx \; x^{2m} \; exp(-\alpha x^2)  =  \frac{(2m-1)!! \sqrt{\pi}}{2^m \alpha^{m} \sqrt{\alpha}}
\end{equation}
We obtain the final result
\begin{equation}
\frac{\langle \phi^{2n} \rangle}{\left( \frac{M^3 t}{3H} \right)^{2n}} = 1+\sum_{k=1}^{n} \frac{(2n)!}{(2k)! (2n-2k)!} \frac{(2k-1)!!}{(2 \alpha \bar{\phi}^2)^k}
\end{equation}
for even powers, and similarly for odd powers we obtain the formula
\begin{equation}
\frac{\langle \phi^{2n+1} \rangle}{\left( \frac{M^3 t}{3H} \right)^{2n+1}} = 1+\sum_{k=1}^{n} \frac{(2n+1)!}{(2k)! (2n-2k+1)!} \frac{(2k-1)!!}{(2 \alpha \bar{\phi}^2)^k}
\end{equation}
where we have defined $\alpha^{-1}=2 \Delta t$ and $\bar{\phi}=(M^3t)/(3H)$, and it is straightforward to check that for
$n=1,2$ one obtains the previous expressions for $\langle \phi^2 \rangle$, $\langle \phi^3 \rangle$ and $\langle \phi^4 \rangle$.

The last two expressions are the main result of this article. As most of the equations in realistic models are solved either numerically or approximately,
it is always desirable to have exact analytical solutions. We find it remarkable that the linear scalar potential
i) can be derived in the framework of well-established quantum field theory, ii) is still in agreement with observations, and iii) the corresponding
Fokker-Planck equation can be recast into the diffusion equation and thus be solved exactly.

A final remark is in order here. Throughout this work we imagine that the scalar field is a spectator to inflation, and we have worked 
perturbatively around a non-dynamical de Sitter background, which is an excellent approximation to inflation. However, the scalar field can no longer
be treated as a spectator when its energy density becomes comparable to the cosmological constant. This happens for a number of e-folds $N_*=H \: t$ given by
\begin{equation}
N_* = \frac{9 H^4 M_p^2}{M^6}
\end{equation}
where $M_p$ is the reduced Planck mass. If we require that this happens after $N=60$ we obtain the following bound on $M$
\begin{equation}
M \leq \left( \frac{9 H^4 M_p^2}{60} \right)^{1/6}
\end{equation}

Before finishing we show that it is possible to obtain the same results following another approach by solving the
Langevin's equation directly. With the initial condition $\phi(0)=0$, the solution for $\phi(t)$ is given by
\begin{equation}
\phi(t)=\frac{M^3 t}{3H}+\int_0^t ds \: f(s)
\end{equation}
From this one can immediately see that $\langle \phi \rangle=(M^3 t)/(3H)$ since $\langle f(t) \rangle=0$. Then,
by squaring the solution and using the property of the stochastic source, namely $\langle f(t_1) f(t_2) \rangle=\Delta \delta(t_1-t_2)$, one 
obtains the previous expression for the second moment. In a similar way the third and the fourth moments can also be computed.

\section{Conclusions}

In the present article we have applied Starobinsky's technique of stochastic inflation to the case of a minimally coupled scalar field with a 
linear self-interaction potential. This type of inflaton potential, although still in agreement with
the latest Planck results, could not be obtained from a fundamental theory of particle physics. However, it recently became relevant since it has been shown
that the linear potential can be obtained in the context of well-defined quantum field theory from a Coleman-Weinberg potential, provided that 
a nonminimal coupling to gravity is also present. We have obtained analytical expressions for the stochastic expectation values $\phi^{2n}$ (even powers)
and $\phi^{2n+1}$ (odd powers) in two ways. First by solving exactly the corresponding Fokker-Planck equation for the density probability function, and 
then directly by using the Klein-Gordon-Langevin equation and the properties of the stochastic source. The two approaches give us the same results as expected.


\section{Acknowledgments}
The author thanks the Funda\c c\~ao para a Ci\^encia e Tecnologia (FCT), Portugal, for the financial support to the 
Multidisciplinary Center for Astrophysics (CENTRA),  Instituto Superior T\'ecnico,  Universidade de Lisboa,  through 
the Grant No. UID/FIS/00099/2013. In addition, he thanks the anonymous reviewer for her/his comments and suggestions 
that improved the quality of the manuscript, and R. Woodard for useful correspondence.


\end{document}